\documentclass{PoS}
\usepackage{amsmath,bm}

\def\be{\begin{equation}}
\def\ee{\end{equation}}
\def\beqn{\begin{eqnarray}}
\def\eeqn{\end{eqnarray}}

\def\ba{\begin{array}{c}}
\def\bat{\begin{array}{cc}}
\def\ea{\end{array}}
\def\bi{\begin{itemize}}
\def\ei{\end{itemize}}

\def\cL{{\cal L}}

\def\cO{{\cal O}}

\newcommand{\eqn}[1]{(\ref{#1})}
\newcommand{\bel}[1]{\be\label{#1}}

\newcommand{\lsim}{~{}_{\textstyle\sim}^{\textstyle <}~}
\newcommand{\rms}{\rm\scriptsize}

\title{Rare kaon decays and CP violation}

\ShortTitle{Rare K decays}

\author{\speaker{Antonio Pich} 
\\
IFIC, Universitat de Val\`encia -- CSIC \\
Parc Cient\'{\i}fic,Catedr\'atico Jos\'e Beltr\'an 2, 
E-46980 Paterna, Spain\\
        E-mail: \email{Antonio.Pich@ific.uv.es}}

\abstract{Owing to the strong suppression of flavour-changing neutral-current transitions in the Standard Model, rare kaon decays constitute a superb tool to constrain hypothetical new-physics interactions. At the same time, they provide many interesting tests of the Standard Model itself, being sensitive both to short-distance electroweak scales and to the long-distance QCD dynamics. A brief overview of the current status is presented. The Standard Model prediction for the direct CP-violating ratio $\varepsilon'/\varepsilon$ is also discussed.}

\FullConference{18th International Conference on B-Physics at Frontier Machines - Beauty2019 -\\
		29 September / 4 October, 2019\\
		Ljubljana, Slovenia}

\begin{document}

\section{Introduction}

The investigation of kaon decays has uncovered many fundamental ingredients of the electroweak theory, such  as flavour quantum numbers,  meson-antimeson mixing, parity  violation, CP violation, quark mixing, and the GIM mechanism \cite{Cirigliano:2011ny}. The kaon decay amplitudes provide very interesting tests of the Standard Model (SM), since they involve an intricate interplay between weak, electromagnetic and strong interactions. Moreover, rare kaon decays are sensitive to short-distance scales ($c$, $t$, $W^\pm$, $Z$) and have the potential to unravel new physics (NP) beyond the SM. In particular, high-precision searches for lepton-flavour violation (LFV)  beyond the $10^{-10}$ level
[$\mathrm{Br}(K_L\to e^\pm\mu^\mp)< 4.7\cdot 10^{-12}$ \cite{Ambrose:1998us}, $\mathrm{Br}(K_L\to e^\pm e^\pm\mu^\mp\mu^\mp)< 4.12\cdot 10^{-11}$ \cite{AlaviHarati:2002eh},
$\mathrm{Br}(K^+\to\pi^+\mu^+ e^-)< 1.3\cdot 10^{-11}$ \cite{Sher:2005sp}, $\mathrm{Br}(K^+\to\pi^+\mu^- e^+)< 5.2\cdot 10^{-10}$ \cite{Appel:2000tc} (90\% CL)]
are actually exploring energy scales above the 10~TeV region.
In addition, the mechanism of CP violation can be accurately tested, both in the observed kaon decay modes and through still undetected processes such as $K_L\to\pi^0 \nu\bar\nu$.

The fundamental flavour-changing transitions among the constituent quarks are characterized by the electroweak scale, but the corresponding hadronic amplitudes are governed by the long-distance behaviour of the strong interactions, {\it i.e.}, the confinement regime of QCD.
In kaon decays, the presence of widely separated mass scales ($m_\pi < m_K \ll M_W$) amplifies the QCD corrections with logarithms of the large  mass ratios.
Using the operator product expansion (OPE) and the renormalization group to integrate out the heavy fields all the way down from $M_W$ to scales $\mu < m_c$, one gets an  effective Lagrangian, defined in the three-flavour theory~ \cite{Gilman:1979bc},
\bel{eq:sd_hamiltonian}
\cL_{\mbox{\rms eff}}^{\Delta S=1} \; = \; -\frac{G_F}{\sqrt{2}}\,
V_{ud}^{\phantom{*}} V_{us}^*\;
\sum_i\, C_i(\mu)\, Q_i \, , 
\qquad\qquad
C_i(\mu)\; =\; z_i(\mu) -  y_i(\mu)\, \frac{V_{td}^{\phantom{*}}V_{ts}^*}{V_{ud}^{\phantom{*}} V_{us}^*}\, ,
\ee
which contains local four-fermion operators $Q_i$,
constructed with the light degrees of freedom ($u$, $d$, $s$; $e$, $\mu$, $\nu_\ell$), modulated by Wilson coefficients $C_i(\mu)$ that are functions of the heavy ($Z$, $W$, $t$, $b$, $c$, $\tau$) masses and encode the short-distance logarithmic corrections. 
The violations of the CP symmetry originate in the $y_i(\mu)$ components, which are proportional to the top-quark mixing factors.

The coefficients $C_i(\mu)$ are currently known at the next-to-leading order (NLO) \cite{Buras:1991jm,Buras:1992tc,Buras:1992zv,Ciuchini:1993vr}, which includes all corrections of $\cO(\alpha_s^n t^n)$ and $\cO(\alpha_s^{n+1} t^n)$, with $t\equiv\log{(M_1/M_2)}$ the logarithm of any ratio of
heavy mass scales ($M_{1,2}\geq\mu$). The long-distance contributions from scales below the renormalization scale $\mu$ are contained in the non-perturbative matrix elements of the operators $Q_i$  between the initial and final hadronic states. These contributions should cancel exactly the renormalization scale (and scheme) dependence of the Wilson coefficients. Unfortunately, a rigorous analytic evaluation of the hadronic matrix elements, keeping full control of the QCD renormalization conventions, remains still a challenging task.

\begin{figure}[t]
\begin{minipage}[c]{.5\linewidth}\centering
\setlength{\unitlength}{0.46mm}          
\begin{picture}(163,133)
\put(0,0){\makebox(163,133){}}
\thicklines
\put(8,124){\makebox(25,10){Energy}}
\put(43,124){\makebox(42,10){Fields}}
\put(101,124){\makebox(52,10){Effective Theory}}
\put(5,123){\line(1,0){153}} {
\put(8,86){\makebox(25,30){$M_W$}}
\put(43,86){\framebox(42,30){\fontsize{10}{12}\selectfont $\ba W, Z, \gamma, g \\
     \tau, \mu, e, \nu_i \\ t, b, c, s, d, u \ea $}}
\put(101,86){\makebox(52,30){Standard Model}}

\put(8,43){\makebox(25,20){$\lsim m_c$}}
\put(43,43){\framebox(42,20){\fontsize{10}{12}\selectfont $\ba  \gamma, g  \, ;\, \mu ,  e, \nu_i \\ s, d, u \ea $}}
\put(101,43){\makebox(52,20){$\cL_{\mathrm{QCD}}^{N_f=3}$,
             $\cL_{\mathrm{eff}}^{\Delta S=1,2}$}}

\put(8,0){\makebox(25,20){$m_K$}}
\put(43,0){\framebox(42,20){\fontsize{10}{12}\selectfont $\ba\gamma \; ;\; \mu , e, \nu_i  \\
            \pi, K,\eta  \ea $}}
\put(101,0){\makebox(52,20){$\chi$PT}}
\linethickness{0.3mm}
\put(64,39){\vector(0,-1){15}}
\put(64,82){\vector(0,-1){15}}
\put(69,72){OPE}
\put(69,29){$N_C\to\infty $}}                     
\end{picture}
\vskip -.3cm\mbox{}
\caption{Evolution from $M_W$ to $m_K$. 
  \label{fig:eff_th}}
\end{minipage}
\hfill
\begin{minipage}[c]{.43\linewidth}\centering
\mbox{}\vskip -.5cm
\includegraphics[width=6.2cm]{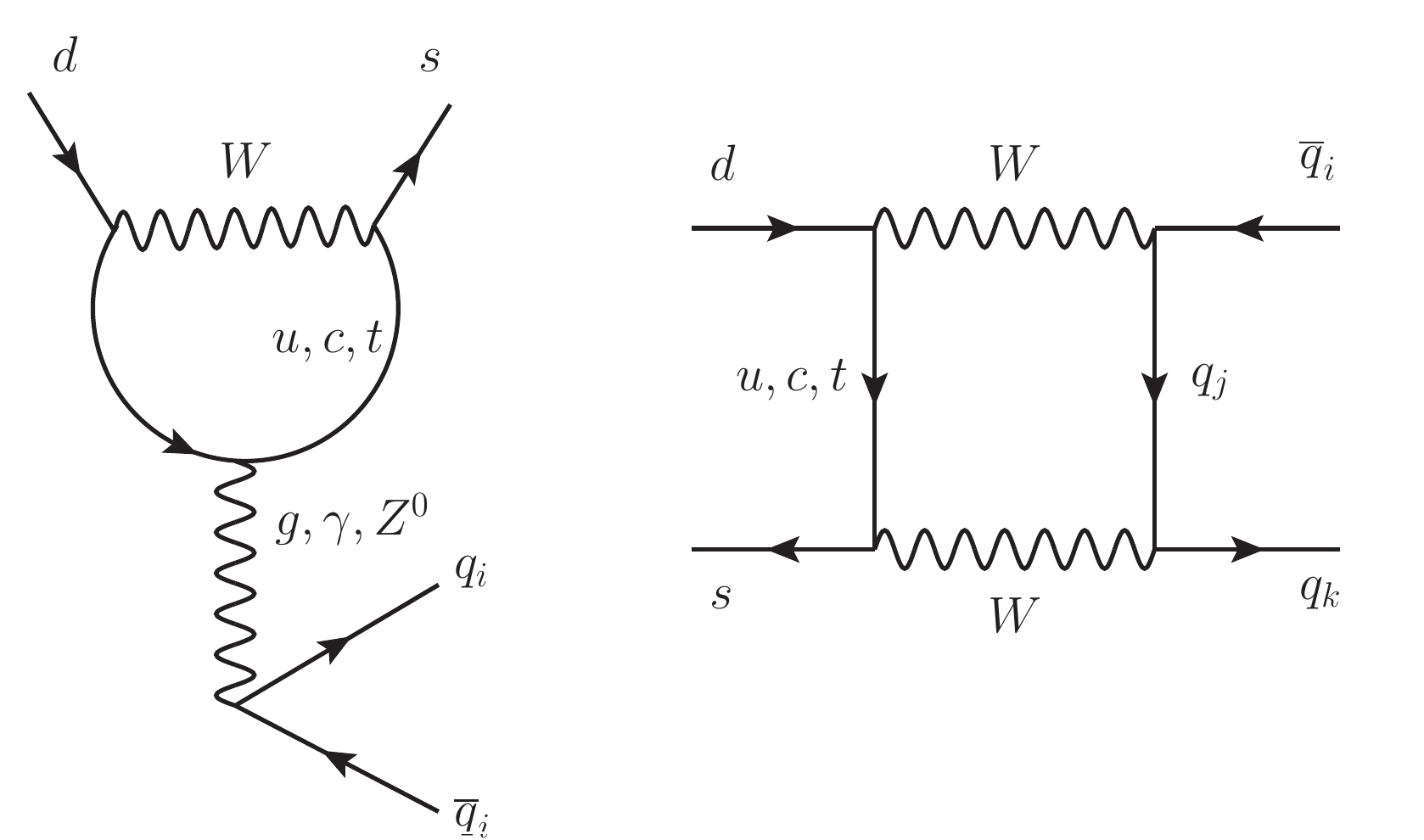}
\caption{\label{fig:sd-diagrams}
Short-distance diagrams.} 
\vskip .3cm
\includegraphics[width=6.5cm]{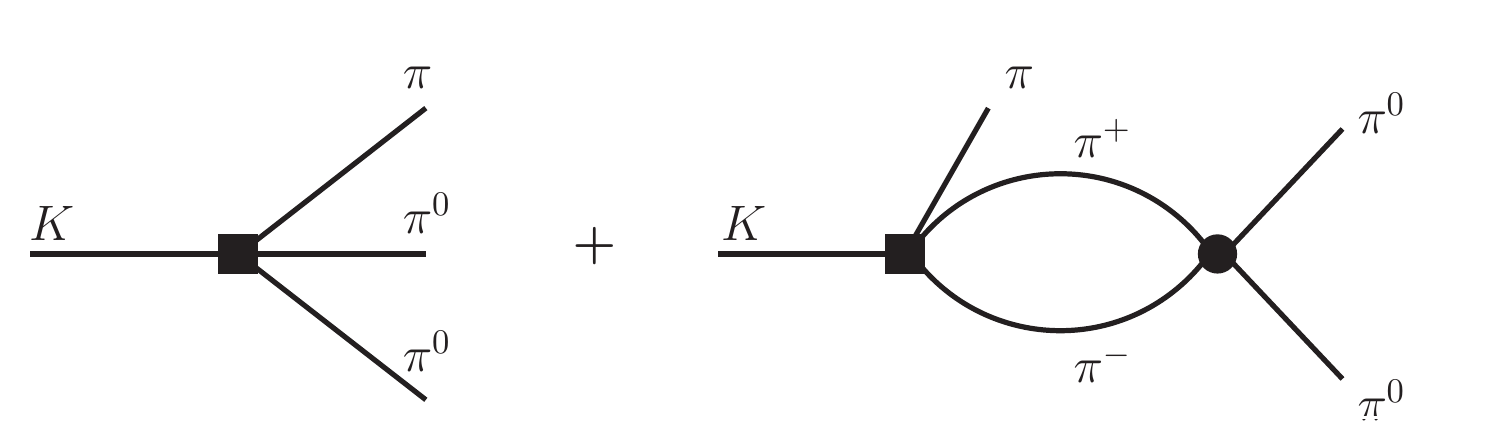}
\caption{\label{fig:ld-diagrams}
Long-distance diagrams.}
\end{minipage}
\end{figure}


At low energies, one can use symmetry considerations to define another effective field theory in terms of the QCD Goldstone bosons ($\pi, K, \eta$). Chiral Perturbation Theory ($\chi$PT) \cite{Weinberg:1978kz,Gasser:1984gg}
describes the  dynamics of the pseudoscalar octet through a perturbative expansion in powers of momenta and quark masses over the chiral symmetry-breaking scale $\Lambda_\chi\sim 1$~GeV.
Chiral symmetry determines the allowed operators, while all short-distance information is encoded in their low-energy couplings (LECs) \cite{Ecker:1994gg,Pich:1995bw}.
At LO the most general effective Lagrangian, with the same
$SU(3)_L\otimes SU(3)_R$ transformation properties as the short-distance
Lagrangian \eqn{eq:sd_hamiltonian}, contains three terms \cite{Cirigliano:2011ny}:
\bel{eq:lg8_g27}
\cL_2^{\Delta S=1} = -{G_F \over \sqrt{2}}  V_{ud}^{\phantom{*}} V_{us}^*
\left\{ g_8  \,\langle\lambda L_{\mu} L^{\mu}\rangle   +
g_{27} \left( L_{\mu 23} L^\mu_{11} + {2\over 3} L_{\mu 21} L^\mu_{13}\right) +
e^2 g_8  g_{\rms ew} F^6\, \langle\lambda U^\dagger Q U\rangle
\right\} ,
\ee
where the $SU(3)$ matrix field
$U\equiv \exp(i \vec{\lambda} \vec{\phi} /F)$ parametrizes the octet of Goldstone bosons,
$L_{\mu}=i F^2 U^\dagger D_\mu U$  represents the 
$V-A$ currents, $\lambda\equiv (\lambda_6 - i \lambda_7)/2$ projects onto the
$\bar s\to \bar d$ transition, 
$Q={1\over 3} \,\mbox{\rm diag}(2,-1,-1)$ is the quark charge matrix
and $\langle\, \rangle$ denotes a 3-dimensional flavour trace.
The $\cO(p^2)$ LECs $g_8$ and $g_{27}$ measure the strength of the two
parts of $\cL_{\mbox{\rms eff}}^{\Delta S=1}$ transforming as
$(8_L,1_R)$ and $(27_L,1_R)$, respectively, under chiral rotations, while the $g_8 g_{\rms ew}$ term is of  $\cO(e^2p^0)$ and
accounts for the $(8_L,8_R)$ piece induced by the electromagnetic penguin operators.

The $\chi$PT framework determines the most general form of the
$K$ decay amplitudes, compatible with chiral symmetry,
in terms of the LECs multiplying the relevant chiral operators.
A first-principle calculation of these LECs would require to perform a non-perturbative matching between $\chi$PT and the underlying SM. While some lattice information is already available in the strong sector, the LECs
can be determined phenomenologically and/or calculated
in the limit of a large number of QCD colours $N_C$. 
Fig.~\ref{fig:eff_th} shows schematically the procedure used
to evolve down from the electroweak scale, where the underlying flavour-changing processes take place (Fig.~\ref{fig:sd-diagrams}), to $m_K$. The short-distance logarithmic corrections $\log{(M/\mu)}$ are summed up with the OPE and the resulting effective Lagrangian $\cL_{\mbox{\rms eff}}^{\Delta S=1}$ is then matched into the low-energy $\chi$PT formalism.

At NLO in the $\chi$PT expansion, one must consider tree-level contributions from additional operators of $\cO(p^4)$ and $\cO(e^2p^2)$, with their corresponding LECs, and quantum loops with the LO Lagrangian \eqn{eq:lg8_g27}. These chiral loops (Fig.~\ref{fig:ld-diagrams}) generate non-polynomial contributions, with logarithms and threshold factors as required by unitarity. The loop corrections contain large infrared logarithms, $\log{(\mu/m_\pi)}$, 
and Goldstone re-scattering contributions  
(final-state interactions) that play a very important role in the kaon decay dynamics.

\section{The CP-violating ratio $\boldsymbol{\varepsilon'/\varepsilon}$}

A tiny difference between the CP-violating ratios $\eta_{nm}\equiv \mathcal{M}[K_L^0\to \pi^n\pi^m]/ \mathcal{M}[K_S^0\to \pi^n\pi^m]\approx\varepsilon \approx 2.2\times 10^{-3}\, \mathrm{e}^{i\pi/4}$, where $nm=+-,00$ denote the final pion charges, was first measured by the CERN NA31 experiment \cite{Burkhardt:1988yh} and later confirmed at the $7.2\sigma$ level with the full data samples of NA31, NA48 and the Fermilab experiments E731 and KTeV \cite{Tanabashi:2018oca}:
\begin{equation}\label{eq:epsp}
\mathrm{Re}(\varepsilon'/\varepsilon) = \frac{1}{3}\,\left(
1 - \left|\frac{\eta_{00}}{\eta_{+-}}\right|^2\right) = (16.6\pm 2.3)\times 10^{-4}\, .
\end{equation}
This important measurement established the presence of direct CP violation
in the decay amplitudes, confirming that CP violation is associated with a $\Delta S=1$ transition as predicted by the SM. 

The first NLO theoretical predictions gave values of $\varepsilon'/\varepsilon$ one order of magnitude smaller than (\ref{eq:epsp}), but it was soon realised that they were missing the important role of the final pion dynamics \cite{Pallante:1999qf,Pallante:2000hk,Pallante:2001he}. Once long-distance contributions are properly taken into account, the theoretical SM prediction turns out to be in good agreement with the experimental value, although the uncertainties are unfortunately large \cite{Gisbert:2017vvj,Cirigliano:2019cpi}:
\begin{equation}
\mathrm{Re}(\varepsilon'/\varepsilon)_{\mathrm{SM}} = (14\pm 5)\times 10^{-4}\, .
\end{equation}

The underlying physics can be easily understood from the kaon data themselves. Owing to Bose symmetry, the two pions in the final state must be in a $I=0$ or $I=2$ configuration. In the absence of QCD corrections, the corresponding $K\to\pi\pi$ decay amplitudes $\mathcal{A}_I\equiv A_I\,\mathrm{e}^{i\delta_I}$
are predicted to differ only by a $\sqrt{2}$ factor. However, their measured ratio is 16 times larger than that (a truly spectacular enhancement generated by the strong forces):
\begin{equation}
\omega\equiv\mathrm{Re}(A_2)/\mathrm{Re}(A_0) \approx 1/22\, ,
\qquad\qquad\quad
\delta_0-\delta_2\approx 45^\circ\, .
\end{equation}
Moreover, they exhibit a huge phase-shift difference that manifests the relevance of final-state interactions and, therefore, the presence of large absorptive contributions to the $K\to\pi\pi$ amplitudes, specially to the isoscalar one. Writing $\mathcal{A}_I = 
\mathrm{Dis} (\mathcal{A}_I) + i\, \mathrm{Abs} (\mathcal{A}_I)$
and neglecting the small CP-odd components, the measured $\pi\pi$ scattering phase shifts at $\sqrt{s}=m_K$ imply that 
\begin{equation}\label{epsp_abs}
\mathrm{Abs}(\mathcal{A}_0)/\mathrm{Dis}(\mathcal{A}_0) \approx 0.82\, ,
\qquad\qquad\quad 
\mathrm{Abs}(\mathcal{A}_2)/\mathrm{Dis}(\mathcal{A}_2) \approx - 0.15\, .
\end{equation}

The direct CP-violating effect involves the interference between the two isospin amplitudes,
\begin{equation}\label{epsp_th}
\mathrm{Re}(\varepsilon'/\varepsilon)\, =\, -\frac{\omega}{\sqrt{2}\, |\varepsilon|}\,\left[\frac{\mathrm{Im} A_0}{\mathrm{Re} A_0} -
\frac{\mathrm{Im} A_2}{\mathrm{Re} A_2}\right]
\, =\,
 -\frac{\omega_+}{\sqrt{2}\, |\varepsilon|}\,\left[\frac{\mathrm{Im} A_0^{(0)}}{\mathrm{Re} A_0^{(0)}}\,\left( 1 -\Omega_{\mathrm{eff}}\right) -
\frac{\mathrm{Im} A_2^{\mathrm{emp}}}{\mathrm{Re} A_2^{(0)}}\right] .
\end{equation}
It is suppressed by the small ratio $\omega$ and, moreover, it is very sensitive to isospin-breaking (IB) corrections \cite{Ecker:1999kr,Cirigliano:2003nn,Cirigliano:2003gt}, parametrized by $\Omega_{\mathrm{eff}}=0.11\pm0.09$ \cite{Cirigliano:2019cpi}, because small IB corrections to 
$A_0$ feed into the small amplitude $A_2$ enhanced by the large factor $1/\omega$. In the right-hand side of Eq.~(\ref{epsp_th}), the $(0)$ superscript indicates the isospin limit, $\omega_+ = \mathrm{Re}(A_2^+)/\mathrm{Re}(A_0)$ is directly extracted from $K^+\to\pi^+\pi^0$ and $A_2^{\mathrm{emp}}$ contains the electromagnetic-penguin contribution to $A_2$ (the remaining contributions are included in 
$\Omega_{\mathrm{eff}}$).

Claims of too small SM values for $\varepsilon'/\varepsilon$ usually originate from perturbative calculations that are unable to generate the physical phase shifts, {\it i.e.}, they predict $\delta_I = 0$ and, therefore, $\mathrm{Abs} (\mathcal{A}_I) = 0$, failing completely to understand the empirical ratios (\ref{epsp_abs}). This unitarity pitfall implies also incorrect predictions for the dispersive components, since they are related by analyticity with the absorptive parts: a large absorptive contribution generates a large dispersive correction that is obviously missed in those calculations. This perturbative problem is more severe in $\varepsilon'/\varepsilon$ because Eq.~(\ref{epsp_th}) involves a delicate numerical balance among the three contributing terms, and naive predictions sit precisely on a nearly-exact cancellation (a 40\% positive correction to the first term enhances the whole result by one order of magnitude).

The $\varepsilon'/\varepsilon$ anomaly was recently resurrected by the lattice RBC-UKQCD collaboration that reported 
$\mathrm{Re}(\varepsilon'/\varepsilon) = (1.38\pm 5.15\pm 4.59)\times 10^{-4}$ \cite{Bai:2015nea,Blum:2015ywa}. The uncertainties are still large, but the quite low central value implies a $2.1\sigma$ deviation from the experimental measurement. This has triggered a revival of the old naive estimates \cite{Buras:2015xba,Buras:2016fys}, some of them making also use of the lattice data \cite{Buras:2015yba,Kitahara:2016nld}, and a large amount of NP explanations (a list of references is given in Refs.~\cite{Gisbert:2017vvj,Cirigliano:2019cpi}). However, it is premature to derive physics implications from the current lattice simulations, since they are still unable to reproduce the known phase shifts. While the lattice determination of $\delta_2$ is only $1\sigma$ away from its physical value, $\delta_0$ disagrees with the experimental result by $2.9\sigma$, a much larger discrepancy that the one quoted for $\varepsilon'/\varepsilon$. Obviously, nobody suggests a NP contribution to the $\pi\pi$ elastic scattering phase shifts. 
The RBC-UKQCD collaboration is actively working in order to improve the present situation.

\section{Rare kaon decays in the SM}

Kaon decays mediated by flavour-changing neutral currents (FCNCs) are strongly suppressed in the SM and, therefore, are very sensitive to NP effects. In the SM, most of these processes are dominated by long-distance contributions, 
making quite challenging their precise theoretical understanding. However, there are also decays governed by short-distance amplitudes, such as $K \rightarrow\pi \nu \bar{\nu}$.

\subsection{$\mathbf{K^0}\boldsymbol{\to\gamma\gamma}$ \ and \ $\mathbf{K^0}\boldsymbol{\to\ell}^{\boldsymbol{+}}\boldsymbol{\ell}^{\boldsymbol{-}}$
}

\begin{figure}[t]
\begin{minipage}[c]{.46\linewidth}\centering
\includegraphics[width=6.2cm]{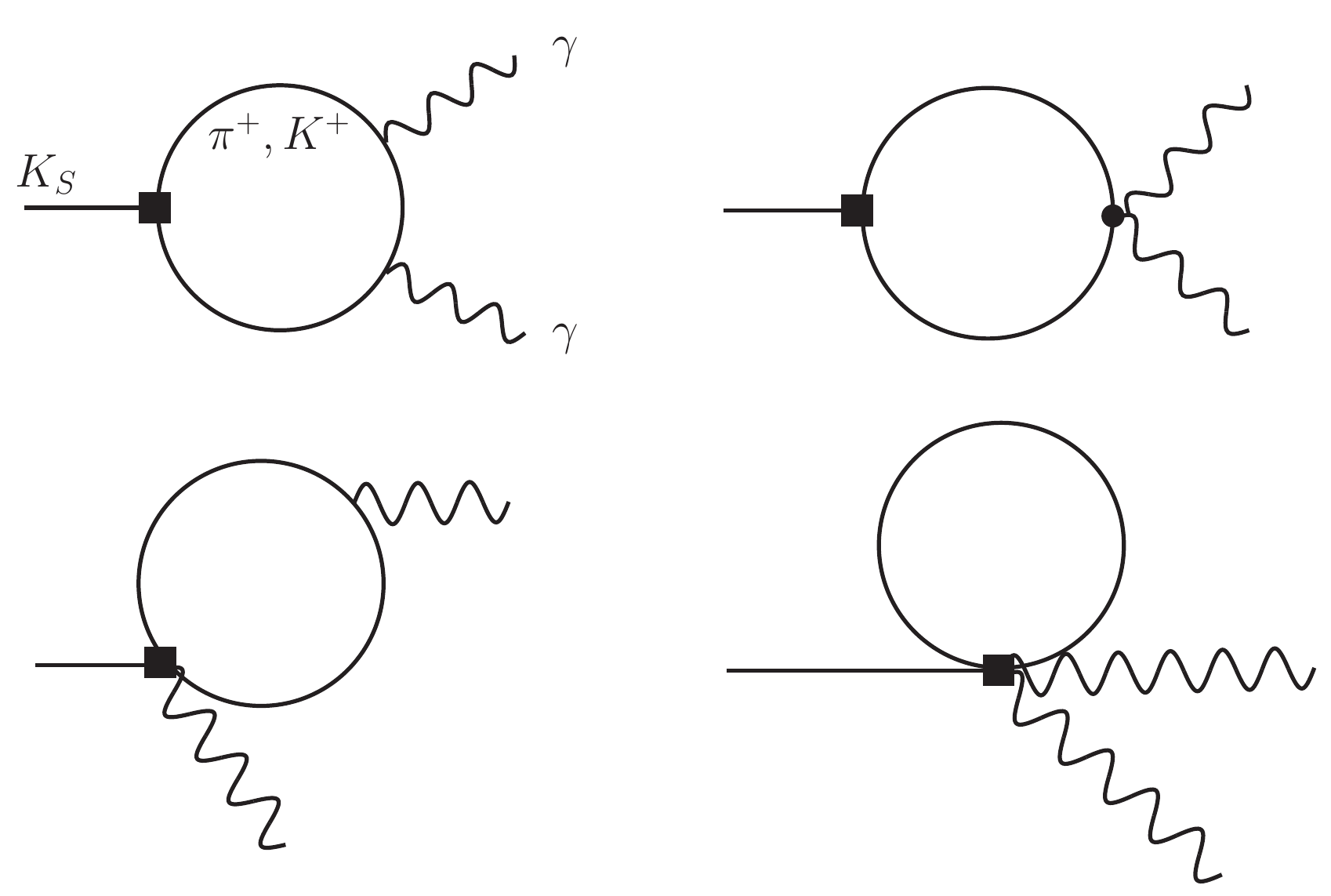}
\caption{Lowest-order 
contributions to $K_S\to\gamma\gamma$.
  \label{fig:KSgg}}
\end{minipage}
\hfill
\begin{minipage}[c]{.46\linewidth}\centering
\mbox{}\vskip -.5cm
\includegraphics[width=6.2cm]{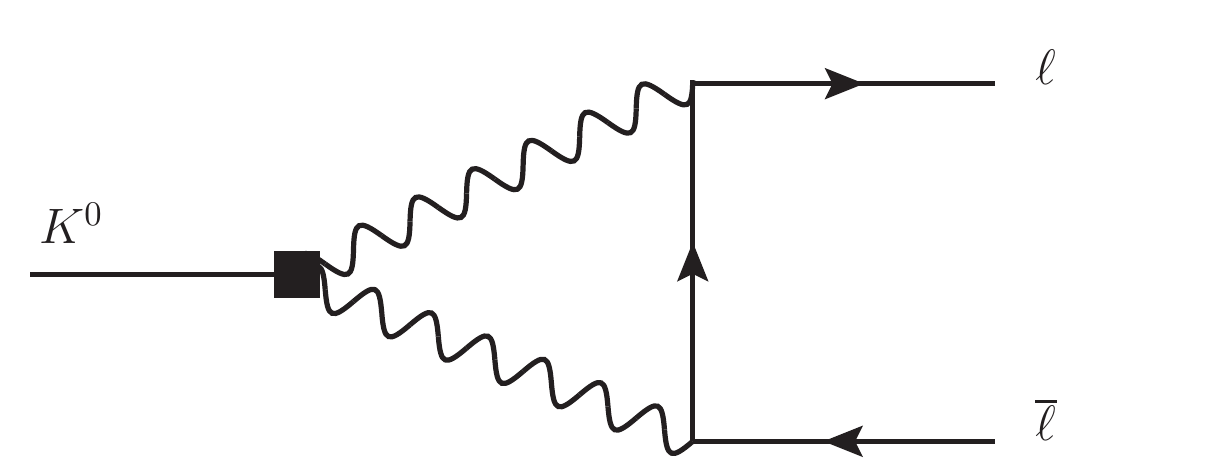}
\caption{\label{fig:KSll}
$2\gamma$ contribution to $K^0\to\ell^+\ell^-$.}
\end{minipage}
\end{figure}

At $\cO(p^4)$ in the $\chi$PT expansion, the symmetry constraints do not allow any local $K_1^0\gamma\gamma$ vertex ($K^0_{1,2}$ denote the CP-even and CP-odd neutral kaon states).
The  decay $K_S^0\to\gamma\gamma$ proceeds then through a one-loop amplitude, with virtual $\pi^+\pi^-$ or $K^+K^-$ pairs (Fig.~\ref{fig:KSgg}),
which is necessarily finite because there are no counterterms to renormalize divergences.
The resulting prediction,
$\mbox{\rm Br}(K_S^0\to\gamma\gamma) = 2.0 \times 10^{-6}$ \cite{DAmbrosio:1986zin,Goity:1986sr},
is slightly lower than the experimental value
$\mbox{\rm Br}(K_S^0\to\gamma\gamma)  = (2.63 \pm 0.17) \times 10^{-6}$ \cite{Tanabashi:2018oca}. 
Full agreement is obtained at
$\cO(p^6)$, once  rescattering corrections ($K_S^0\to\pi\pi\to\pi^+\pi^-\to\gamma\gamma$) are included
 \cite{Kambor:1993tv}.


The 2-loop amplitude $K_S^0 \rightarrow \gamma^* \gamma^* \rightarrow \ell^+
\ell^-$ (Fig.~\ref{fig:KSll}) is also finite  \cite{Ecker:1991ru}
because chiral symmetry forbids any CP-invariant local contribution at this order.
The predicted rates,
$\mathrm{Br}(K_S^0 \rightarrow e^+ e^-) = 2.1\times 10^{-14}$
and
$\mathrm{Br}(K_S^0 \rightarrow \mu^+ \mu^-) = 5.1\times 10^{-12}$  \cite{Ecker:1991ru},
are well below the experimental upper bounds
$\mathrm{Br}(K_S^0 \rightarrow e^+ e^-) < 9\times 10^{-9}$   \cite{Ambrosino:2008zi}
and
$\mathrm{Br}(K_S^0 \rightarrow\mu^+ \mu^-) < 2.1\times 10^{-10}$ \cite{Aaij:2017tia,Aaij:2020sbt}
(90\%  CL). 
This calculation allows us to compute
the longitudinal polarization $P_L$ of either muon in the decay
$K_L^0\rightarrow \mu^+ \mu^-$, a CP-violating observable which
in the SM is dominated by indirect CP violation from
$K^0$--$\bar K^0$ mixing. One finds
$|P_L| = (2.6\pm 0.4)\times 10^{-3}$ \cite{Ecker:1991ru}.

\subsection{$\mathbf{K}\boldsymbol{\to\pi\gamma\gamma}$}

\begin{figure}[t]
\begin{minipage}[c]{.46\linewidth}\centering
\includegraphics[width=6.9cm]{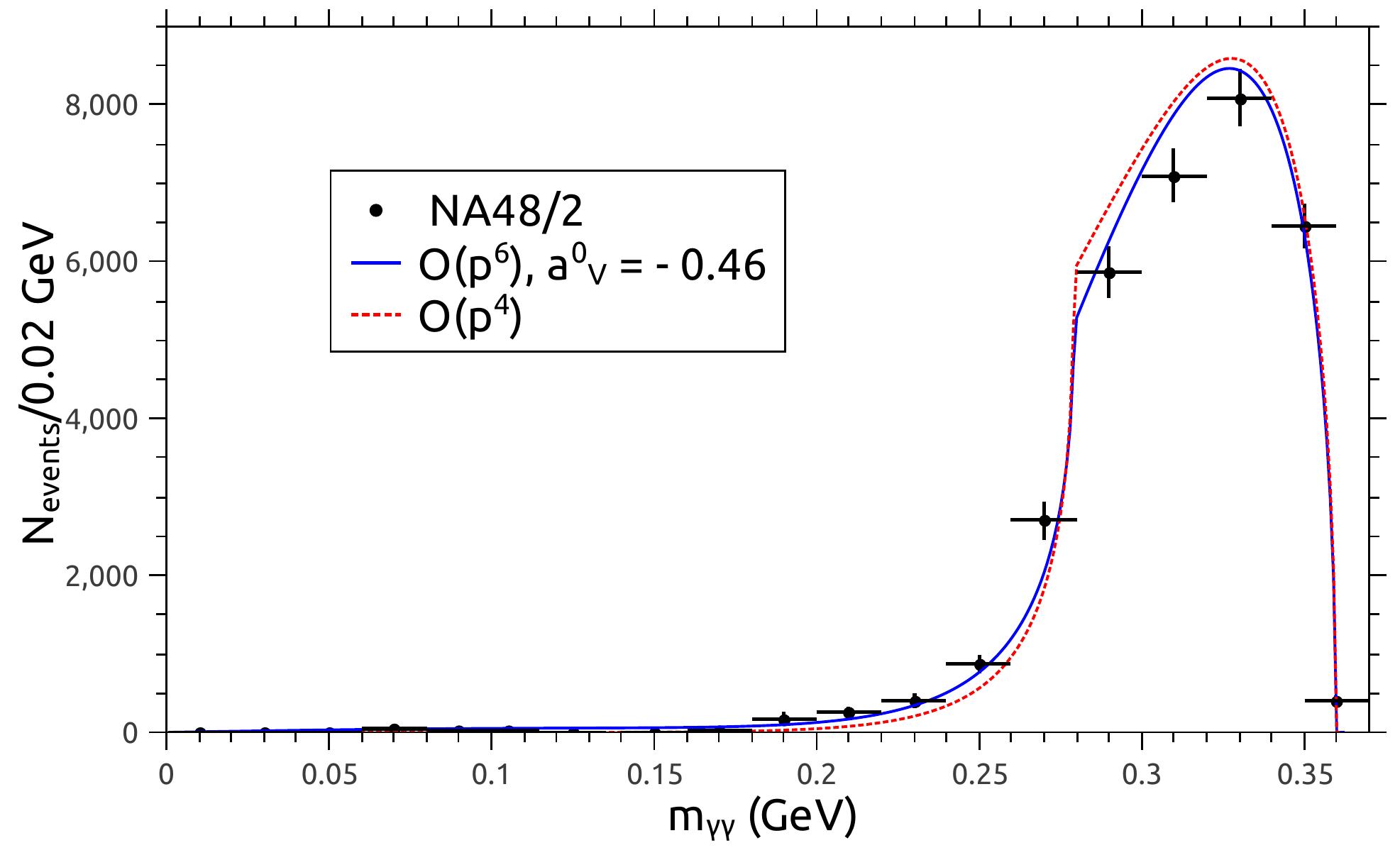}
\caption{$K_L\to\pi^0\gamma\gamma$ spectra
at $\cO(p^4)$ and $\cO(p^6)$ in
$\chi$PT. The data are from Ref.~\cite{Lai:2002kf}.
  \label{fig:KL_pgg}}
\end{minipage}
\hfill
\begin{minipage}[c]{.45\linewidth}\centering
\mbox{}\vskip -.2cm
\includegraphics[width=6.cm]{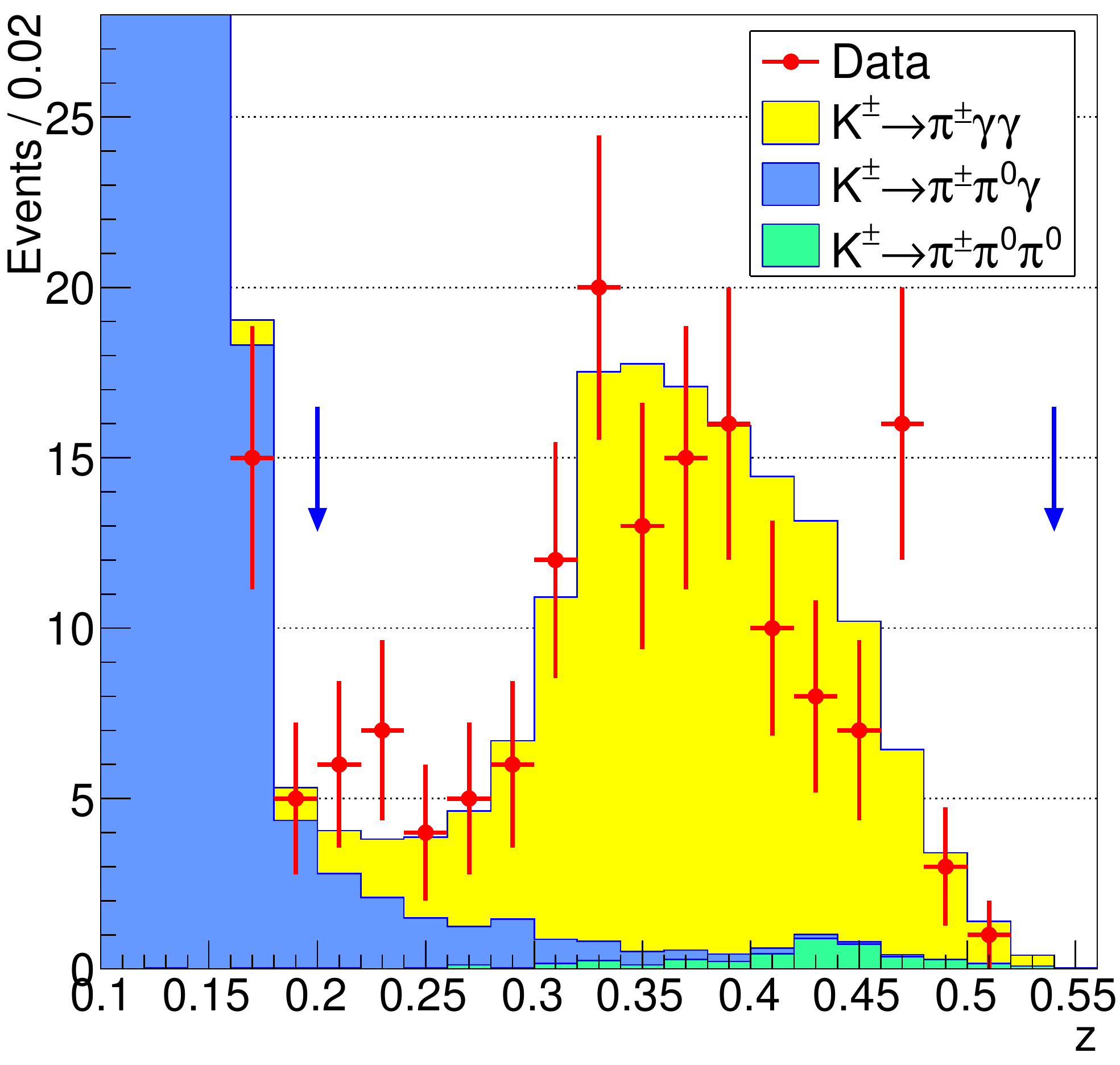}
\vskip -.2cm
\caption{\label{fig:K+p+gg}
Measured $K^\pm\to\pi^\pm\gamma\gamma$ spectrum and estimated signal from an $\cO(p^6)$ $\chi$PT fit \cite{Batley:2013daa}.}
\end{minipage}
\end{figure}

Again, the symmetry constraints do not allow any tree-level contribution to $K_2^0\to\pi^0\gamma\gamma$ from $\cO(p^4)$ terms in the $\chi$PT Lagrangian. The decay amplitude is therefore determined by a finite
loop calculation \cite{Ecker:1987fm,Cappiello:1988yg,Sehgal:1989pw}.
Due to the large absorptive $\pi^+\pi^-$ contribution, the spectrum in
the invariant mass of the two photons is predicted to have a very characteristic behaviour
(dotted line in Fig.~\ref{fig:KL_pgg}), peaked at high values of $m_{\gamma\gamma}$.
The agreement with the measured distribution \cite{Lai:2002kf} is remarkably good. However, the $\cO(p^4)$ prediction for
the rate,  $\mathrm{Br}(K_L \rightarrow \pi^0 \gamma \gamma) = 6.8 \times 10^{-7}$ \cite{Ecker:1987fm}, is
significantly smaller than the present PDG average,
$\mathrm{Br}(K_L^0 \rightarrow \pi^0 \gamma \gamma) = (1.27 \pm 0.03) \times 10^{-6}$ \cite{Tanabashi:2018oca},
indicating that higher-order corrections are sizeable.
Unitarity corrections from $K_L^0 \rightarrow \pi^+ \pi^- \pi^0$
\cite{Cohen:1993ta,Cappiello:1992kk} and local vector-exchange contributions \cite{Cohen:1993ta,Ecker:1990in}
restore the agreement at ${\cal O}(p^6)$.

A quite similar spectrum is predicted \cite{Ecker:1987hd} for the charged mode $K^\pm\to\pi^\pm\gamma\gamma$,
but in this case there is a free LEC already at $\cO(p^4)$.
Corrections of $\cO(p^6)$ have been also investigated \cite{DAmbrosio:1996cak}.
Both the measured spectrum and the rate can be correctly reproduced \cite{Batley:2013daa}, as illustrated in Fig.~\ref{fig:K+p+gg}.

\subsection{$\mathbf{K_L^0\boldsymbol{\to\pi}^0 e^{\boldsymbol{+}} e^{\boldsymbol{-}}}$}

This decay 
is an interesting process in looking for new CP-violating signatures, because
$K_2^0\to\pi^0\gamma^*$ violates CP \cite{Ecker:1987hd,Donoghue:1994yt}.
The CP-conserving amplitude proceeds through a $2\gamma$ intermediate state and
is suppressed by an additional power of $\alpha$. Using the $K_L^0 \rightarrow \pi^0 \gamma\gamma$ data,
the CP-conserving rate is found to be below $10^{-12}$ \cite{Cirigliano:2011ny}.
The $K_L^0 \rightarrow \pi^0 e^+ e^-$ transition is then dominated by the $\cO(\alpha)$ CP-violating contributions \cite{Ecker:1987hd},
both from $K^0$--$\bar K^0$ mixing and direct CP violation. The estimated rate,
$\mathrm{Br}(K_L \rightarrow \pi^0 e^+ e^-) =  (3.1  \pm 0.9)\times 10^{-11}$ \cite{Cirigliano:2011ny,Buras:1994qa,Buchalla:2003sj},
is only a factor 10 smaller than the present (90\% CL) upper bound of $2.8 \times 10^{-10}$ \cite{AlaviHarati:2003mr}
and should be reachable in the near future.

\subsection{$\mathbf{K\boldsymbol{\to\pi\nu\bar\nu}}$}

Long-distance effects play a negligible role in $K^+\to\pi^+\nu\bar\nu$
and $K_L^0\to\pi^0\nu\bar\nu$.
These processes are dominated by short-distance loops ($Z$ penguin,
$W$ box), involving the heavy top quark. The $K^+$ decay mode receives also sizeable contributions from internal charm-quark exchanges.
The decay amplitudes are proportional to the hadronic matrix element of the $\Delta S=1$ vector current,
which (assuming isospin symmetry) can be obtained from $K_{\ell 3}$ decays:
\bel{eq:pnn}
T(K\to\pi\nu\bar\nu)\,\sim\, \sum_{i=c,t}
F(V_{id}^{\phantom{*}} V_{is}^*;x_i)\;
\left(\bar\nu_L\gamma_\mu\nu_L\right)\;
\langle\pi |\bar s_L\gamma^\mu d_L|K\rangle
\, ,
\qquad\qquad x_i\equiv m_i^2/M_W^2 \, .
\ee
The small long-distance and isospin-violating corrections can be estimated within $\chi$PT.
The $K_L^0\to\pi^0\nu\bar\nu$ transition violates CP and is completely dominated by direct CP violation, the
contribution from $K^0$--$\bar K^0$ mixing being only of the order of 1\%.
Taking the CKM inputs from global fits, one predicts
$\mathrm{Br} (K_L^0 \to \pi^0 \nu \bar{\nu}) =  (2.9 \pm 0.3) \times 10^{-11}$
and
$\mathrm{Br} (K^+  \to \pi^+ \nu \bar{\nu}) = (8.5 \pm 0.6) \times 10^{-11}$
\cite{Buras:2005gr,Brod:2010hi,Gorbahn_kaon2019}.
The uncertainties are largely parametrical,
due to CKM input, $m_{c}$, $m_{t}$  and $\alpha_{s} (M_Z)$.

The current (90\% CL) upper bounds on the charged \cite{NA62_kaon2019}
and neutral \cite{Ahn:2018mvc} modes are
\be\label{eq:Kpnn_exp}
\mathrm{Br} (K^+  \to \pi^+ \nu \bar{\nu}) < 1.85 \times 10^{-10}\, ,
\qquad\qquad
\mathrm{Br} (K_L^0\to \pi^0 \nu \bar{\nu}) <   3.0  \times 10^{-9}\, .
\ee
The ongoing CERN NA62 experiment aims to reach $\mathcal{O}(100)$ $K^+  \to \pi^+ \nu \bar{\nu}$ events (assuming SM rates), while increased sensitivities on the $K_L^0 \to \pi^0 \nu \bar{\nu}$ mode are expected to be achieved by the KOTO experiment at J-PARC.

\section{Constraints on scalar leptoquarks from rare kaon decays}

Rare kaon decays put strong constraints on NP interactions with non-trivial flavour dynamics. As an illustration, and motivated by the flavour anomalies reported recently in $B$ decays \cite{Pich:2019pzg}, let us consider the implications of kaon data on generic couplings of hypothetical scalar leptoquarks (LQs) to the SM fermions~\cite{Mandal:2019gff}:
\begin{equation}
\label{eq:Lfull}
\mathcal{L}_{\mathrm{LQ}}\, =\, 
\overline{Q^c}\, i \tau_2\, y_{\tiny S_1} L\; S_1 
+ \overline{d^c_R}  \, y_{\tiny \tilde{S_1}} \ell_R\; \tilde{S}_1 
+\overline \ell_R \,y_{\tiny R_2}\, R_2^\dagger\, Q 
- \overline{d}_R  \, y_{\tiny \tilde{R_2}}\,  \tilde{R}_2^T\, i \tau_2 L
+ \overline{Q^c}\, y_{\tiny S_3} \,i\tau_2  \, {\boldsymbol \tau\bf \cdot S_3} \, L
+ \mathrm{h.c.} \, .
\end{equation}
We have included the five possible types of scalar LQs, coupling to SM particles, with the following
$SU(3)_C\otimes SU(2)_L\otimes U(1)_Y$ quantum numbers~\cite{Davidson:1993qk,Dorsner:2016wpm}: $S_1\,({\bf \bar{3},\,1},\,1/3)$, $\tilde{S}_1\,({\bf \bar{3},\,1},\,4/3)$, $R_2~({\bf 3,\,2},\,7/6)$, $\tilde{R}_2\,({\bf 3,\,2},\,1/6)$ and $S_3\,({\bf\bar{3},\,3},\,1/3)$.
$Q$ and $L$ are the left-handed quark and lepton doublets, $d_R$ and $\ell_R$ the corresponding right-handed singlets and
$f^c\equiv \mathcal{C}\bar f^{\, T}$ indicates the charge-conjugated field of the fermion $f$. All fermion fields carry flavour indices and $y_{LQ}$ are arbitrary Yukawa matrices in flavour space. Eq.~(\ref{eq:Lfull}) only displays those couplings relevant for kaon decays.

\begin{figure}[t]
	\begin{center}
		\includegraphics[width=0.42\linewidth]{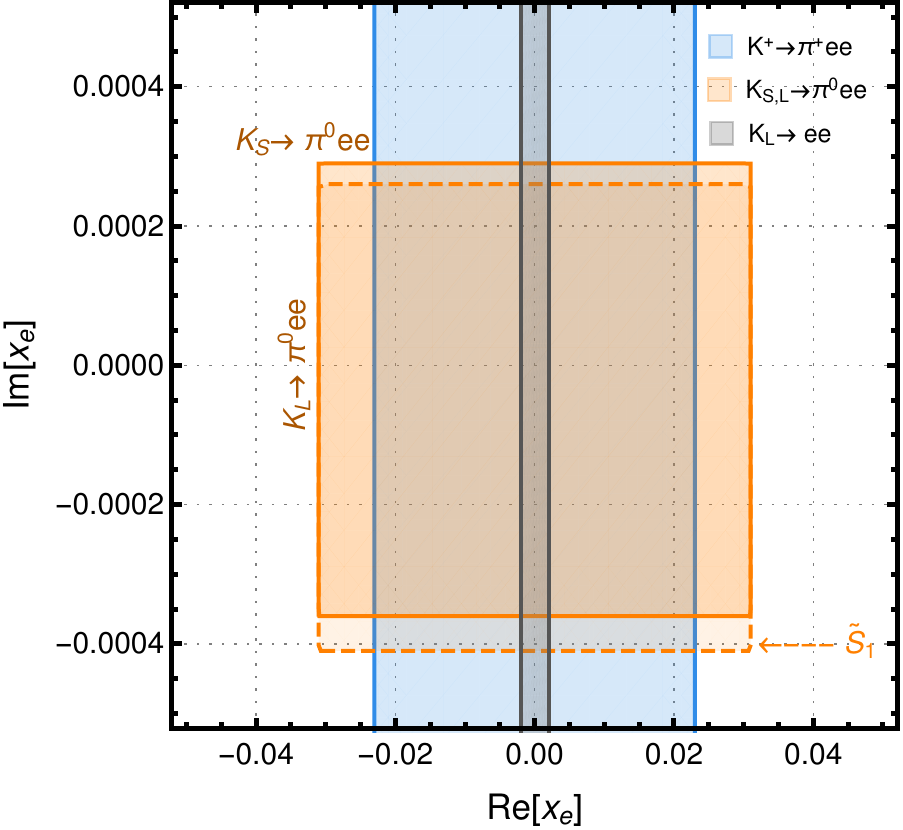} \hskip 50pt
		\includegraphics[width=0.42\linewidth]{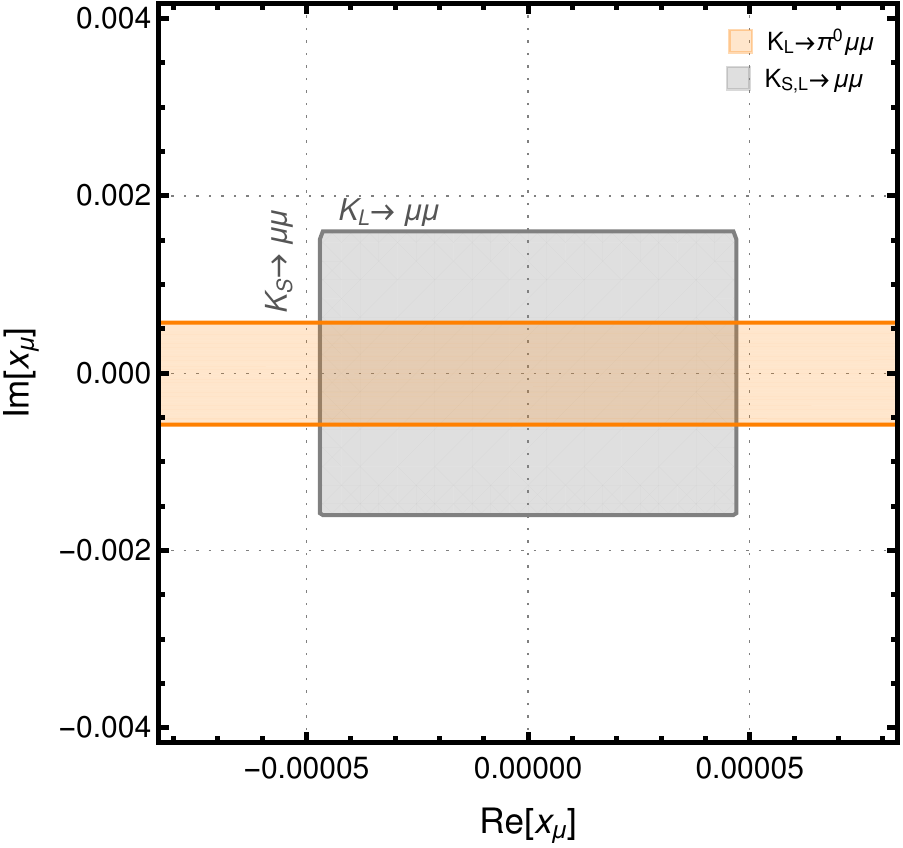}
		\caption{Allowed regions in the plane $(\mathrm{Re}[x_\ell], \mathrm{Im}[x_\ell])$,
arising from leptonic and rare semileptonic kaon decays, for the electron (left panel) and muon (right panel) channels \cite{Mandal:2019gff}. }\label{fig:Kaon}
	\end{center}
\end{figure}

The exchange of a heavy LQ between two fermionic currents induces tree-level  contributions to the FCNC transitions $K^0\to\ell^+\ell^-$ and $K\to\pi\ell^+\ell^-$. They are governed by the following combinations of LQ parameters:
\begin{equation}
x_e = \left(\frac{1~\mathrm{TeV}}{M_{\mathrm{LQ}}}\right)^2\times\left\{
\begin{array}{c}
y_{\mathrm{LQ}}^{11} \; ( y_{\mathrm{LQ}}^{12})^* \\[2pt] y_{\mathrm{LQ}}^{11} \; (y_{\mathrm{LQ}}^{21})^*
\end{array} \right.\, ,
\qquad\qquad\qquad
x_\mu = \left(\frac{1~\mathrm{TeV}}{M_{\mathrm{LQ}}}\right)^2\times\left\{
\begin{array}{c}
y_{\mathrm{LQ}}^{21}\; (y_{\mathrm{LQ}}^{22})^* 
\\[2pt] 
y_{\mathrm{LQ}}^{12}\; (y_{\mathrm{LQ}}^{22})^*
\end{array} \right.\, ,
\end{equation}
where $x_e$ and $x_\mu$ correspond to the electron and muon modes, respectively. The first line in the brackets corresponds to $R_2$, while the second line refers to $\tilde R_2$, $\tilde S_1$ and $(4\times)\, S_3$ (the LQ $S_1$ does not contribute at tree level to these processes).
The current constraints on $x_\ell$ from different kaon decay modes are displayed in Fig.~\ref{fig:Kaon} for the electron (left panel) and muon (right panel) final states. The $K_S^0$ and $K_L^0$ decays are complementary, providing separate access to both the real and imaginary parts of the NP couplings, while the decays of the charged kaon restrict their absolute value. The strongest constraints come from $K_L\to\mu^+\mu^-$ [Re$(x_\mu)$], $K_L\to\pi^0\mu^+\mu-$ [Im$(x_\mu)$], $K_L\to\pi^0 e^+ e^-$ [Im$(x_e)$]  and $K_L\to e^+ e^-$ [Re$(x_e)$].

More stringent constraints can be derived from the current experimental limits on LFV transitions. The 90\% CL upper bounds 
$\mathrm{Br}(K_L\to e^\pm \mu^\mp) < 4.7\times 10^{-12}$  \cite{Ambrose:1998us} and 
$\mathrm{Br}(K_L\to \pi^0 e^\pm \mu^\mp) < 7.6\times 10^{-11}$ \cite{Abouzaid:2007aa} imply \cite{Mandal:2019gff}
\begin{eqnarray}
\left(\frac{1~\mathrm{TeV}}{M_{\mathrm{LQ}}}\right)^2\times\left\{
\begin{array}{c}
\left| y_{\mathrm{LQ}}^{21} \; ( y_{\mathrm{LQ}}^{12})^* 
+ y_{\mathrm{LQ}}^{22} \; ( y_{\mathrm{LQ}}^{11})^* \right|
\\[2pt] 
\left| y_{\mathrm{LQ}}^{21} \; ( y_{\mathrm{LQ}}^{12})^* 
+ y_{\mathrm{LQ}}^{11} \; ( y_{\mathrm{LQ}}^{22})^* \right|
\end{array} \right. &<\; & 1.9\times 10^{-5}\, ,
\\[5pt]
\left(\frac{1~\mathrm{TeV}}{M_{\mathrm{LQ}}}\right)^2\times\left\{
\begin{array}{c}
\left| y_{\mathrm{LQ}}^{21} \; ( y_{\mathrm{LQ}}^{12})^* 
- y_{\mathrm{LQ}}^{22} \; ( y_{\mathrm{LQ}}^{11})^* \right|
\\[2pt] 
\left| y_{\mathrm{LQ}}^{21} \; ( y_{\mathrm{LQ}}^{12})^* 
- y_{\mathrm{LQ}}^{11} \; ( y_{\mathrm{LQ}}^{22})^* \right|
\end{array} \right. &<\; & 2.9\times 10^{-4}\, ,
\end{eqnarray}
respectively, while $\mathrm{Br}(K^+\to \pi^+\mu^+e^-) < 1.3 \times 10^{-11}$ \cite{Sher:2005sp} leads to
\begin{equation}
\left(\frac{1~\mathrm{TeV}}{M_{\mathrm{LQ}}}\right)^2\times\;\left(
\left| y_{\mathrm{LQ}}^{21} \; ( y_{\mathrm{LQ}}^{12})^* \right|\, , \,
\left| y_{\mathrm{LQ}}^{11} \; ( y_{\mathrm{LQ}}^{22})^* \right|
\right)\; <\;
1.9\times 10^{-4}
\end{equation}
for the four LQ types.

\begin{figure}[th]
	\begin{center}
		\includegraphics[width=0.45\linewidth]{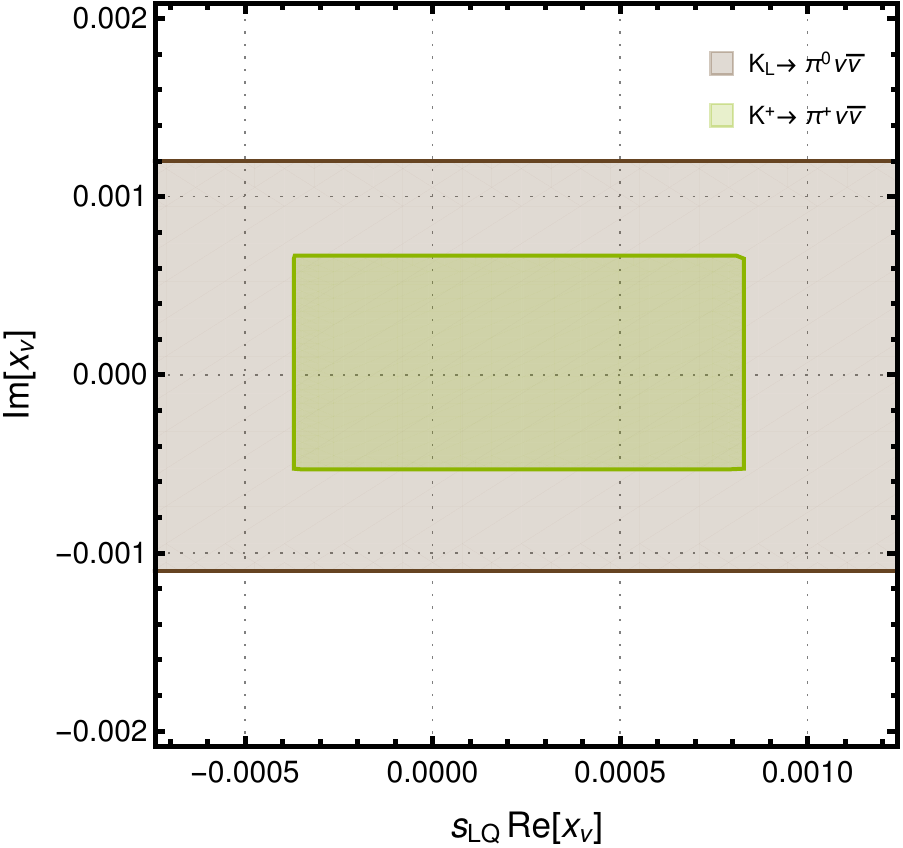}
		\caption{Allowed regions in the plane $(s_{\mbox{\tiny LQ}}\,\mathrm{Re}[x_\nu], \mathrm{Im}[x_\nu])$,
arising from $K\to\pi\nu\bar\nu$ decays \cite{Mandal:2019gff}. The sign factor $s_{\mbox{\tiny LQ}}=+1$ for $S_{1,3}$, while $s_{\mbox{\tiny LQ}}=-1$ for $\tilde{R}_2$.}\label{fig:KaonNu}
	\end{center}
\end{figure}

The $K\to\pi\nu\bar\nu$ decay modes only receive tree-level contributions from  $S_1$, $S_3$ and $\tilde R_2$. For identical neutrino flavours $\nu_\ell\bar\nu_\ell$ in the final state, the corresponding constraints are shown in Fig.~\ref{fig:KaonNu} \cite{Mandal:2019gff}. The relevant combination of LQ parameters is in this case
\begin{equation}
x_\nu = \left(\frac{1~\mathrm{TeV}}{M_{\mathrm{LQ}}}\right)^2\times
\hat y_{\mathrm{LQ}}^{1\ell}\, (\hat y_{\mathrm{LQ}}^{2\ell})^* \, ,
\end{equation}
where $\hat y_{\mathrm{LQ}} = y_{\mathrm{LQ}}\, U$ with $U$ the PMNS neutrino mixing matrix. Notice that the neutral mode only constrains $\mathrm{Im}(x_\nu)$, while $K^+\to\pi^+\nu\bar\nu$ puts limits on both the real and imaginary parts of $x_\nu$.

The three LQs induce also $K\to\pi\nu_m\bar\nu_n$ decay modes with different neutrino flavours, which should not evade the experimental limits in Eq.~(\ref{eq:Kpnn_exp}). The $K^+$ decay implies the upper bound  \cite{Mandal:2019gff}
%
\begin{equation}\label{eq:Kpnn_difFlav_Kp}
\left(\frac{1~\mathrm{TeV}}{M_{\mathrm{LQ}}}\right)^2\times\;
\big[\!\sum \limits_{m\not=n}\!|\hat{y}_{\mathrm{LQ}}^{1m}\, 
({\hat{y}}_{\mathrm{LQ}}^{2n})^*|^2\,\big]^{1/2}\; < \; 6.0	\times 10^{-4}\, ,
\end{equation}
while the neutral decay mode puts the constraint \cite{Mandal:2019gff}
\begin{equation}\label{eq:Kpnn_difFlav_K0}
\left(\frac{1~\mathrm{TeV}}{M_{\mathrm{LQ}}}\right)^2\times\;
\big[\!\sum \limits_{m\not=n}\!|
		\hat{y}_{\mathrm{LQ}}^{1m}\, ({\hat{y}}_{\mathrm{LQ}}^{2n})^* - 
		\hat{y}_{\mathrm{LQ}}^{2m}\, ({\hat{y}}_{\mathrm{LQ}}^{1n})^*
	|^2\,\big]^{1/2} \; <\; 1.1\times 10^{-3}\, .
\end{equation}

The KOTO collaboration has recently reported the observation of four $K_L^0 \to\pi^0  \nu \bar{\nu}$ events, with an expected background
of only $0.05\pm 0.02$ events~\cite{KOTO_kaon2019}. Removing one of the events that is suspected to originate in underestimated upstream activity background, the quoted single event sensitivity of $6.9\times 10^{-10}$ would correspond to 
$\mathrm{Br}(K_L^0 \to\pi^0  \nu \bar{\nu}) \sim 2\times 10^{-9}$, well above the new Grossman-Nir limit~\cite{Grossman:1997sk} implied by the NA62 upper bound on $\mathrm{Br}(K^+ \to\pi^+  \nu \bar{\nu})$:
\begin{equation}
\mathrm{Br}(K_L^0 \to\pi^0  \nu \bar{\nu})\; <\; 4.2\times 
\mathrm{Br}(K^+ \to\pi^+  \nu \bar{\nu}) \; <\; 7.8\times 10^{-10}\, .
\end{equation}
This limit is valid under quite generic assumptions, provided the lepton flavour is conserved, and it can be directly inferred from the predicted LQ-induced decay amplitudes \cite{Mandal:2019gff}, if there are only identical neutrino flavours in the final state.

In order to reach the KOTO signal, one needs a sizeable decay amplitude
into neutrinos with different flavours ($n\not= m$). This could be easily achieved within the $S_1$, $S_3$ and $\tilde R_2$ LQ scenarios. A confirmation of the KOTO events would just imply that the combination of LQ couplings in Eq.~(\ref{eq:Kpnn_difFlav_K0}) takes a non-zero value quite close to its current upper bound, indicating a violation of lepton flavour. 
Other possible NP interpretations have been already considered in Refs.~\cite{Kitahara:2019lws,Egana-Ugrinovic:2019wzj,Dev:2019hho,Fabbrichesi:2019bmo,Li:2019fhz,Jho:2020jsa, Liu:2020qgx}.

\section*{Acknowledgements}

I want to thank the organizers of Beauty 2019 for the invitation to present this overview. I also thank
V. Cirigliano, H. Gisbert, R. Mandal and A. Rodr\'{\i}guez-S\'anchez for a very productive and enjoyable collaboration.
This work has been supported in part by the Spanish Government and ERDF funds from the EU Commission [grant FPA2017-84445-P] and the Generalitat Valenciana [grant Prometeo/2017/053].

\bibliographystyle{JHEP}
\bibliography{FlavAnom,EpsilonpRefs,kaon} 

%

\end{document}